\documentclass[12pt,preprint]{aastex}

\def\etal   {{\rm ~et al.~}}
\def\kms    {\ifmmode{{\rm ~km~s}^{-1}}\else{~km~s$^{-1}$}\fi}

\shorttitle{Mid-infrared Images of Orion BN/KL}
\shortauthors{Greenhill\etal}

\begin{document}

\title{High Angular Resolution Mid-infrared Imaging of Young Stars in Orion BN/KL}

\author{L. J. Greenhill\altaffilmark{1}, D. Y. Gezari\altaffilmark{2}, W. C. Danchi\altaffilmark{2},
J. Najita\altaffilmark{3}, J. D. Monnier\altaffilmark{4}, and P. G. Tuthill\altaffilmark{5}}

\altaffiltext{1}{Harvard-Smithsonian Center for Astrophysics, 60 Garden Street, Cambridge, MA 02138;
greenhill@cfa.harvard.edu}

\altaffiltext{2}{NASA/Goddard Space Flight Center Code 685, Greenbelt, MD 20771}

\altaffiltext{3}{NOAO, 50 North Cherry Avenue, Tucson, AZ 85719}

\altaffiltext{4}{University of Michigan, 941 Dennison Building, Ann Arbor, MI 48109}

\altaffiltext{5}{School of Physics, Sydney University, NSW 2006, Australia}

\begin{abstract}

We present Keck LWS images of the Orion BN/KL star forming region obtained in the first
multi-wavelength study to have $0\rlap{.}''3$-$0\rlap{.}''5$ resolution from $4.7\mu$m to $22\mu$m. 
The young stellar objects designated infrared source\,{\it n} and radio source\,{\it I} are believed to
dominate the  BN/KL region.  We have detected extended emission from a probable accretion disk around
source\,{\it n} but infer a stellar luminosity on the order of only 2000\,L$_\odot$.  Although source\,{\it
I} is believed to be more luminous, we do not detect an infrared counterpart even at the longest
wavelengths.   However, we resolve the closeby infrared source, IRc2, into an arc of knots $\sim 10^3$\,AU
long at all wavelengths.  Although the physical relation of source\,{\it I} to IRc2 remains ambiguous, we
suggest these sources mark a high density core ($10^7$-$10^8$\,pc$^{-3}$ over $\sim10^3$\,AU) within the
larger BN/KL star forming cluster. The high density  may be a consequence of the core being young and
heavily embedded.  We suggest the energetics of the BN/KL region may be dominated by this
cluster core rather than one or two individual sources.

\end{abstract}


\section{Introduction}

Orion BN/KL is the closest region of high-mass star formation ($\sim 450$\,pc) and though also the
archetype, it is not necessarily well understood \citep[e.g.,][]{genzel89}.  For instance,  the sources
that drive extensive outflow activity in BN/KL \citep[e.g.,][]{wright96}  and contribute most to its
luminosity have not been identified with certainty \citep[e.g.,][]{mr95}.  This is true because BN/KL is
crowded, exhibiting $\sim 20$ compact infrared (IR) peaks distributed over $\sim 20''$ and because 
extinction is high and variable across the region \citep{gezari98}.  As high-mass stars often form
in clusters \citep[][and references therein]{gl99},  there are presumably numerous embedded protostars
and young stellar objects (YSOs) in the region, many of which have probably not been recognized yet.   

Two notable sources in BN/KL exhibit centimeter-wave continuum and as a result are believed to be
luminous YSOs, IR source\,{\it n}  \citep{lonsdale82} and radio source\,{\it I} \citep{churchwell87}. 
Source\,{\it n}, also known as  radio source\,{\it L} \citep{churchwell87}, exhibits a bipolar radio jet
\citep{mr95}, hard X-ray emission \citep{garmire00,feigelson02}, and $\lambda 2.3\mu$m CO overtone
emission probably from the hot surface of an accretion disk \citep{luhman00}.  Source\,{\it I} has been
difficult to study because it has not been detected at IR wavelengths, probably due to high extinction.
It lies close to the mid-IR source IRc2 \citep{dyg92,mr95} for which $A_V\sim 60$ \citep{gezari98} and
the peak of the Orion Hot Core, for which probably $A_V\gg10^2$ \citep{genzel89,migenes89,plambeck95}. 
However, source\,{\it I} excites H$_2$O and SiO maser emission in dense, hot material at radii $<10^3$\,AU
\citep{genzel89,dyg92}, which may be used to study gas dynamics in detail
\citep[e.g.,][]{greenhill98,greenhill03}.

We have imaged BN/KL in the mid-IR at resolutions of $0\rlap{.}''3$-$0\rlap{.}''5$  with the intention to
characterize sources of excitation.  This is the
first multiwavelength study with sub-arcsecond resolution up to $22\mu$m.
\citep[cf.,][]{dougados93,rouan04,shuping04}.  We concentrate on a largely qualitative interpretation of the
data. Quantitative modeling and analysis of the whole region will be discussed in a separate paper.

\section{Observations and Reduction} 

We observed the BN/KL region with the Keck-I telescope on 1999 December 21, 2000 October 14, and 2002
August 20.  We used the Long Wavelength Spectrometer (LWS; Jones \& Puetter 1993) in imaging mode at seven
filter wavelengths from $4.7\mu$m to $22.0\mu$m (Table\,1) to obtain frames of 128$\times$128 pixels ($\sim
10''\times10''$), with which we created mosaics.  We chopped at 5\,Hz and nodded the telescope both using
fields $30''$ west in 1999 and 2002 and $40''$ north in 2000.  Registration of overlapping frames is
uncertain to $\la 1$ pixel.  To register images obtained at different wavelengths, we used the peak of the
Becklin-Neugebauer object (BN) as a reference.  When BN lay outside a mosaic, we used IRc7.  The formal
uncertainty in registration is $< 1$ pixel.   

We used observing procedures and data reduction techniques discussed by \citet{gezari92} and
implemented in the MOSAIC image processing package \citep{vg93}.  Observations of $\alpha$ CMa
demonstrated the high quality of  the point source responses.  Images at
$12.5\mu$m and longer wavelengths are diffraction limited ($0\rlap{.}''3$-$0\rlap{.}''5$), while seeing,
chopping/nodding error, and smoothing during processing limit resolution to $\sim 0\rlap{.}''3$ for shorter
wavelengths.  To calibrate flux densities, we scaled images at each wavelength to match the intensities
for  BN or IRc7 reported by \citet{gezari98}, interpolated to the LWS passbands (Table\,1).  

We calibrated the pixel scale to first order using observations in 1999 of the
binary XY\,Per\, for which we adopted a $1\rlap{.}''33$ separation and $256.3^\circ$ position angle
\citep{bertout99}.  We made second order adjustments to align the IR and radio centroids of BN
and source\,{\it n} on a $\lambda12.5\mu$m mosaic, where we defined the radio
centroid for source\,{\it n} as the  midpoint between the lobes.  The adopted scale and
rotation were $0\rlap{.}''0840$ pixel$^{-1}$ and $2.1^\circ$ clockwise (i.e., north is rotated
counterclockwise from vertical on the images).  Uncertainties of 1-2\% and $0.5^\circ$ 
include the $\sim 0\rlap{.}''02$\,yr$^{-1}$ proper motion for BN \citep{plambeck95}. We rechecked the plate
scale in 2002 with observations of $\alpha$\,Her, for which we adopted a $4\rlap{.}''84$  separation and
$105^\circ$ position angle \citep{mason00,docobo00}, and obtained $0\rlap{.}''0831$ pixel$^{-1}$
and $4^\circ$ clockwise, with uncertainties of 2\% and $1^\circ$, in good agreement with earlier results.

We adopted the 1999 plate scale for images obtained in 2000 because a parallelogram distortion precluded
our estimating a new solution.  However, because the sub-fields shown here are small (containing
just IRc2 and IRc7) the resulting errors are $<2$ pixels.  We estimated this
error empirically by registering the peak of IRc7 observed at $\lambda12.5\mu$m in 2000 and 1999,
applying the registration offsets to images from 2000 for other wavelengths, and measuring the 
misalignment in the peaks of IRc2.  

To register the radio position of source\,{\it I} on the IR images, we assumed the radio and IR centroids
for BN were coincident and used the angular separation of BN and source\,{\it I} estimated by
\citet{mr95}.  For images obtained in 2000, we measured angular offsets from (closeby) IRc7 to minimize
the effect of the parallelogram distrotion.  The total uncertainty in radio-IR registration is 1-2
pixels.  (We note that the proper motion of BN contributes only a $\sim 0.4$ pixel error to the 
registration.)

\section{Results and Discussion}

\subsection{Source\,{\it n} as an intermediate luminosity star}

We have found that emission from source\,{\it n} at $\lambda8.0\mu$m and  $\lambda 11.7\mu$m is elongated
nearly east-west (Figure\,1) and orthogonal to the bipolar radio structure detected previously
by \citet{mr95}. In contrast, other sources in the $\lambda8.0\mu$m and  $\lambda 11.7\mu$m images
are not elongated (e.g., IRc7).  We propose that the elongation arises from warm dust in the accretion
disk inferred previously from the presence of CO overtone emission at $\lambda2.3\mu$m \citep{luhman00}. We
fit a single component elliptical Gaussian model to each image. At $\lambda8.0\mu$m, the source size (half
power full width)is $(0\rlap{.}''74\times0\rlap{.}''50) \pm 0\rlap{.}''08$, and at $\lambda 11.7\mu$m, the
size is $(0\rlap{.}''75 \times 0\rlap{.}''53) \pm 0\rlap{.}''12$ ($\sim 340\times 230$\,AU).  The position
angles  are $115^\circ\pm13^\circ$ and $101^\circ\pm24^\circ$, respectively, while the  position angle of
the bipolar radio structure is $12\pm1^\circ$.   The shape of source\,{\it n} at longer wavelengths  
is more uncertain, because of plateau emission between IRc4 and IRc7.   However, if we assume the
disk emits as a blackbody peaking close to the shortest wavelength at
which it is detected ($8.0\mu$m), then the characteristic temperature is on the order of
300\,K.

Although source\,{\it n} is extended at mid-IR wavelengths, it is point-like in high angular
resolution observations from $\lambda2.2\mu$m to $\lambda5\mu$m \citep{dougados93,stolovy98}.
This near-IR emission may arise from the inner edge of the accretion disk, close to where
dust is sublimated.  For the disks of Herbig AeBe stars that are resolved with interferometry, this inner
radius is $<0.2-10$\,AU \citep{tmd01,mmg02}.  At the distance of BN/KL, the near-IR emission would
subtend $<0\rlap{.}''02$, thus explaining its point-like appearance in observations so far.    

The luminosity of source\,{\it n} is not known.  However, mid-IR emission from an accretion disk may be
expected to arise from surface material heated by  stellar radiation, and we can use the characteristic
radius, R, and temperature, T, to estimate the luminosity of source\,{\it n}, $L_*$.  Dust near the surface
of the disk is optically thick to stellar radiation (mostly optical and ultraviolet) but optically thin to
re-emission (mostly mid-IR).  For thermal equilibrium, $L_* \sim 16\pi R^2 \sigma T^4 (Q_e/Q_a)$,
where $Q_e$ and $Q_a$ are grain emissivities for emission and absorption, and $\sigma$ is the
Stefan-Boltzmann constant.  For absorption of stellar radiation, $Q_a=1$, while for re-emission, $Q_e\sim
2\pi a/\lambda$, where $a$ is a characteristic grain size.  For $\lambda\sim10\mu$m and $a\sim 0.1\mu$m
(typical for interstellar dust), $Q_e/Q_a\sim 0.06$ and $L_*\sim 2000$\,L$_\odot$ (mid B-type).  This
luminosity is consistent with the observation that at $\lambda1.6\mu$m source\,{\it n} is $\sim 10$ times
dimmer than BN (1-2$\times10^4$\,L$_\odot$; Scoville\etal 1983), given that source\,{\it n} is
1.2~magnitudes less red between $\lambda 1.1\mu$m and $1.6\mu$m \citep{luhman00}, if we assume the two
stars have approximately the same intrinsic near-IR color.  Nonetheless, the estimated luminosity should be
regarded with caution. It depends on uncertain values for $a$ and $T$ rather than radiative transfer
modeling of the spectral energy distribution, and the estimated luminosity exceeds the
apparent luminosity of $\sim10$\,L$_\odot$ (Table\,1).  Moreoever, identification of source\,{\it n} as a
hard X-ray source is a concern since a B-type star is unlikely to be active, though the X-rays may be
associated with a less massive companion or the jet.

\subsection{Source\,{\it I}, IRc7, and IRc2}

An IR counterpart to radio source\,{\it I} is long soughtafter.  The LWS observations are the first
to resolve the angular separation between source\,{\it I} and the comparatively bright IRc2 at wavelengths
up to $22\mu$m.  Nonetheless, we do not detect emission from a counterpart with an upper limit on the order
of 1\,Jy at $\lambda 8.0\mu$m and 10\,Jy at $\lambda 22\mu$m.  We suggest that emission from source\,{\it
I} is blocked by the Orion Hot Core.  \citet{plambeck95} estimate the average optical depth of the core is
$\sim 0.05$ at $\lambda3.5$\,mm. The wavelength scaling of opacity to shorter wavelengths is uncertain
because it depends on grain composition and the amount of processing. However, if we assume spherical
silicate dust grains \citep{mathis90}, then the optical depth is $>300$ at $\lambda8.0\mu$m and
$\lambda22\mu$m.  This is sufficient to obscure the mid-IR dust emission from a $10^3$\,K blackbody of
diameter up to 80\,AU and luminosity up to
$\sim6\times10^4$\,L$_\odot$, which is comparable to the luminosity of the BN/KL region \citep{gezari98}.  
It is also sufficiently large to explain nondetection at $\lambda2.2\mu$m in Hubble Space Telecope (HST)
and Keck observations \citep{stolovy98, hc00}.

IRc7 is relatively compact and prominent at all observed wavelengths longer than $8\mu$m (Figures\,2 and
3).  This is probably indicative of star formation at a relatively early phase.  IRc7 also lies close (in
projection) to the radio axis of source\,{\it n} \citep{stolovy98}, and HST observations  at
$\lambda2.2\mu$m have detected a fan of emission opening away from source\,{\it n} with a bright point at
the vertex.   This morphology and orientation may signify interaction with radiation and outflow incident
from source\,{\it n}.  Alternatively the near-IR fan may be a scatting cone along the rotation axis of a
protostellar system embedded in an optically thin envelope or halo.

IRc2 is shaped like a $\sim 10^3$\,AU long arc that opens to the southwest, in the direction of
source\,{\it I}, which lies close to one end (Figure\,2).  The locations of knots along the arc depend
on wavelength and are distributed $0\rlap{.}''6$ to $2 \rlap{.}''7$ from source\,{\it I}.  Similar
emission distributions at $\lambda 3.8, 5,$ and $12.5\mu$m are reported by \citet{dougados93},
\citet{rouan04}, and \citet{shuping04}.  Emission at the shorter wavelengths is the closest to
source\,{\it I} while emission at $\lambda22.0\mu$m is among the farthest (Figure\,3).  Although suggestive
of central heating, emission at wavelengths near the peaks of the silicate absorption features at
$9.7\mu$m and $18.5\mu$m also tends to lie away from source\,{\it I} (Figures\,2 and 3).  If local
extinction toward source\,{\it I} is enhanced, then it is surprising that the short wavelength emission
arises closeby to source\,{\it I}.  At the same time, we note that the shortest wavelength emission yet
detected from IRc2, at $\lambda2.2\mu$m, is associated with the knot ``B'' \citep{dougados93}, $\sim 1''$
from source\,{\it I} \citep{stolovy98, hc00}.

The relationship of IRc2 to source\,{\it I} is ambiguous.  Because massive stars form in clusters, we speculate
that the $\sim 5$ knots in IRc2 may be individual sites of star formation, which together with
source\,{\it I} form at least part of a high density core within the larger ($\sim 20''$) BN/KL star forming
cluster.  In support of this we note that (1) the relative distributions of emission in IRc2 from
$\lambda2.2\mu$m to $\lambda22.0\mu$m do not  exhibit an overall systematic pattern (e.g., central heating),
and (2) knot ``C'' \citep{dougados93} is both closely associated with a cluster of OH masers (Figure\,2)
and  X-ray emission (E. Feigelson, private communication).  (If the X-ray emission is the result of a wind
driven by source\,{\it I}, then other knots would probably be X-ray sources as well.)  The
inferred characteristic density of $10^7$-$10^8$\,pc$^{-3}$ over
$\sim 10^3$\,AU is unusually large and greatly exceeds the density of the closeby Trapezium
\citep{hh98}.   Because high cluster densities are associated with high stellar masses
\citep[][and references therein]{testi99} we might expect to observe precursors to O-type stars, but the
large density may be due to cluster youth. As clusters evolve they expand because of stellar dynamics  and
the expulsion of gas \citep[][and references therein]{kroupa04}.  Clusters that are still deeply embedded
may be expected to exhibit unusually high central densities though observational evidence is weak. 

We cannot yet exclude the possibility that IRc2 is heated externally, by source\,{\it I}. The tip
of IRc2  recedes systematically as wavelength increases from $\lambda5\mu$m to
$\lambda22.0\mu$m (Figure\,2), which is suggestive of a local temperature gradient.  In principle,
IRc2 could mark the edge of a wind-blown cavity.  Relying on maps of SiO masers \citet{greenhill98} propose
that source\,{\it I} drives a conical outflow to the north and west, which IRc2 just happens to subtend. 
\citet{greenhill03} present an alternative model in which source\,{\it I} binds an edge-on flared disk whose
limbs extend north and west from the star.  In this case IRc2 straddles the disk plane.  The mid-IR arc
would comprise irradiated disk surface material and peripheral low latitude material heated by other YSOs.

\section{Conclusions}

We have detected the probable signature of an accretion disk around source\,{\it n} at mid-IR wavelengths,
and we estimate the stellar luminosity is $\sim 2000$\,L$_\odot$.  We have not identified any counterpart
to radio source\,{\it I}, despite higher angular resolution and broader wavelength coverage than previous
studies.  We resolve IRc2 into knots at up to $\lambda22.0\mu$m and propose that they are sites of star
formation.  Together, source\,{\it I} and IRc2 may mark the dense core
($10^7$-$10^8$\,M$_\odot$\,pc$^{-3}$) of the larger BN/KL infrared cluster. 
Many past studies have sought to demonstrate that luminosity and outflow associated with BN/KL are driven
by one or two sources.  This seems increasingly unlikely. We note that source\,{\it n} has only a
modest luminosity and its jet axis is poorly aligned with outflow axes observed on scales of
$10^4$\,AU \citep[e.g.,][]{schultz99}.  Moreover, members of the putative IRc2-source\,{\it I} subcluster
are ready candidates to drive outflow collectively.  In a later paper, we will present a
radiative model of the LWS data, with estimates of dust temperatures, line-of-sight optical depths, and
intrinsic luminosities for the dominant mid-IR sources in BN/KL.

\acknowledgements{We thank E. Feigelson, S. Wolk, and the COUP consortium for sharing X-ray source
coordinates before publication. F. Varosi provided technical support for the MOSAIC software.   The Keck
Observatory was made possible by the financial support of the W. M. Keck  Foundation and is operated as a
scientific partnership among  CalTech, the University of California, and NASA.
}

\clearpage

\begin{deluxetable}{cc@{\extracolsep{-0.07in}}cc@{\extracolsep{-0.15in}}c@{\extracolsep{-0.15in}}c@{\extracolsep{0.15in}}c}
\tablewidth{3.9in}
\tablecaption{Peak Specific Intensities}
\tablehead{
     \colhead{Filter} &
     \colhead{Band} &
     \colhead{Ep.\tablenotemark{(a)}}      &
     \colhead{IRc2} &
     \colhead{IRc7} &
     \colhead{{\it n}} & 
     \colhead{BN} \\
     \colhead{($\mu$m)} &
     \colhead{($\mu$m)} &
     \colhead{} &
     \colhead{} &
     \colhead{(Jy\,arcsec$^{-2}$)\tablenotemark{(b)}} &
     \colhead{} &
     \colhead{}   \\
}

\startdata

4.7 (M)& 4.4-5.0  & 2 & 0.64 & 0.15 & \nodata & 58  \\
8.0    & 7.5-8.2   & 1 & 15   & 5.0  & 4.4    & \nodata \\
10.7   & 10.0-11.4 & 3 & 1.3  & 2.6  & 0.96    & 93 \\
11.7   & 11.2-12.2 & 3 & 4.9  & 7.2  & 2.4   & 140 \\
12.5   & 12.0-13.0 & 1 & 13   & 17   & 3.3    & 152 \\
17.65  & 17.3-18.2 & 3 & 16   & 34   & 2.3    & 175 \\
22.0   & 21.0-23.0 & 3 & 43   & 86   & 3.2    & 190 \\

\enddata

\tablenotetext{(a)}{Epoch for the tabulated photometry (see Section\,2).}

\tablenotetext{(b)}{Intensities are uncertain to $\sim 30\%$. The zero level was taken from a sky
position $\sim 3''$ south of source\,{\it I}.  Plateau emission between IRc4 and
IRc7 has been subtracted in the case of source\,{\it n}.  Images are scaled to match listed IRTF peak
intensities for BN, or IRc7 at $\lambda 8.0\mu$m \citep{gezari98} interpolated to LWS passbands.}


\end{deluxetable}

\clearpage

\begin{figure}
\epsscale{0.4}
\plotone{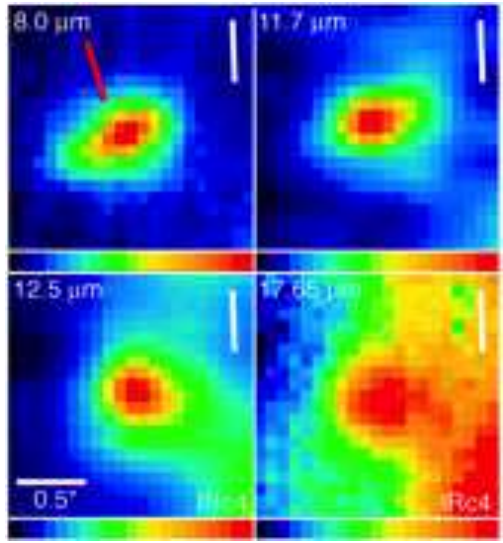}
\caption{Images of source\,{\it n} obtained in 1999.  The elongations at $\lambda8.0$ and
$\lambda11.7\mu$m are within $\sim 10^\circ$ of perpendicular to the jet axis (red bar).   
The peak at each wavelength is listed in Table\,1, and the specific intensity
scale is linear.  The upper right bar in each panel indicates north.}
\end{figure}

\begin{figure}
\epsscale{1.0}
\plotone{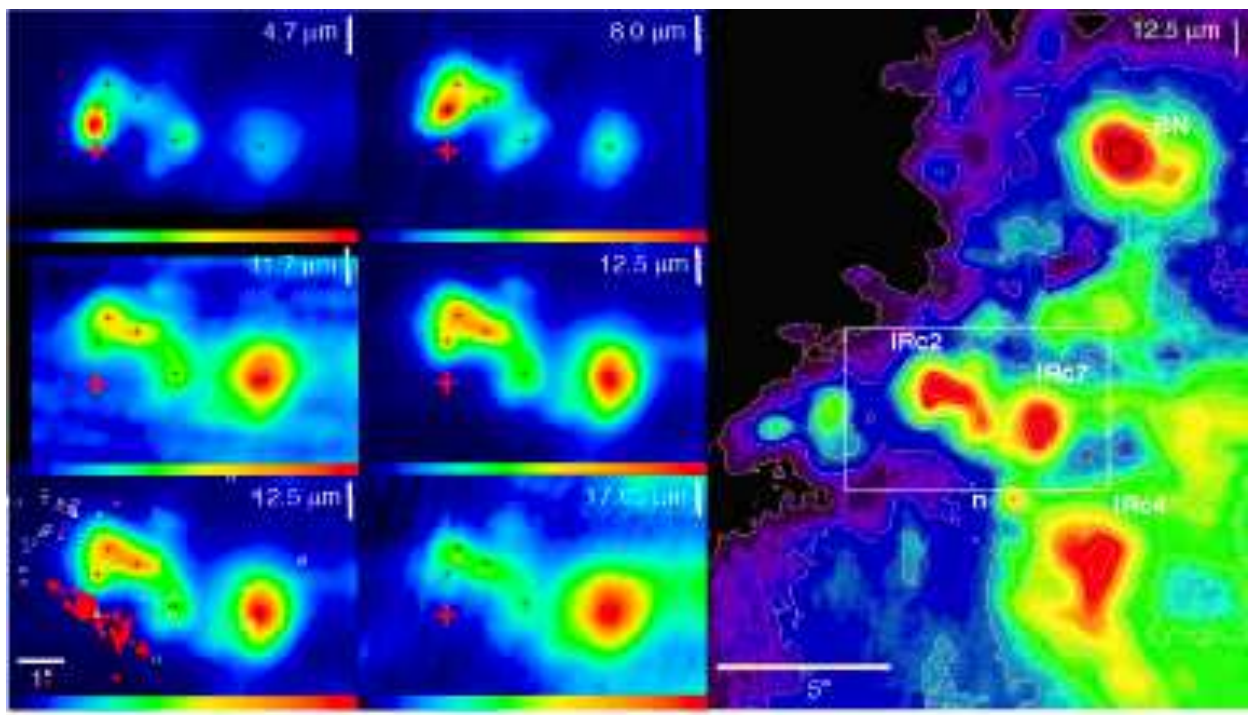}
\caption{Mid-IR images obtained in 1999 and 2000 ($\lambda 4.7\mu$m only). {(\it right)} Mosaic of
$\lambda12.5\mu$m emission.  Contours correspond to 1.2, 1.6, 2.1, 2.9, 4.0, 5.6, 7.8, 11, and 16 Jy
arcsec$^{-2}$.  The white bar indicates north.  The box outlines the region shown in close-up. {\it (left)} 
Close-ups of IRc2 and IRc7.  Black crosses mark the approximate positions of
$\lambda12.5\mu$m peaks and are provided for registration. Symbol size indicates the
alignment uncertainty of the panels.  The red cross marks the
position of source\,{\it I}, and symbol size reflects uncertainty.  The distributions of H$_2$O {\it (red
circles)} and OH masers {\it (white squares)} are also shown \citep{greenhill98,johnston89}.  The H$_2$O
masers lie in a high-density outflow from source\,{\it I} (arrows). The OH masers lie in lower density gas
and are probably pumped by far-IR radiation. The peak specific intensities of IRc2 and IRc7 at each
wavelength are listed in Table\,1, and the specific intensity scale is linear.}
\end{figure}

\begin{figure}
\epsscale{0.4}
\plotone{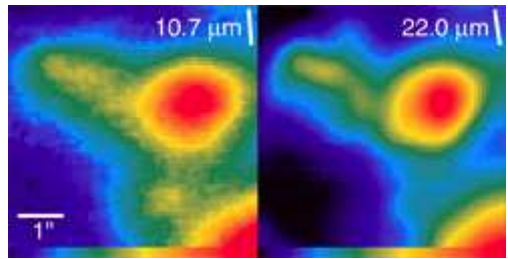}
\caption{Images of IRc2, IRc7, and source\,{\it n} obtained in 2002.  Peak specific intensities are listed
in Table\,1, and the specific intensity scale is linear.  The white bar in each panel indicates north.}
\end{figure}


\begin{thebibliography}{}

\bibitem[Bertout, Robichon, \& Arenou (1999)] {bertout99} Bertout, C., Robichon, N.,\& Arenou, F. 1999,
A\&A, 352, 574

\bibitem[Churchwell\etal (1987)]{churchwell87} Churchwell, E., Felli, M., Wood, D. 
O. S., \& Massi, M. 1987, \apj, 321, 516

\bibitem[Docobo\etal (2000)]{docobo00} Docobo, J. A., Tamazian, V. S., Balega, Y. Y., Blanco, J.,
Maximov, A. F., \& Vasyuk, V. A., 2000, A\&A 366, 868


\bibitem[Dougados\etal (1993)]{dougados93} Dougados, C., Lena, P., Ridgway, S. T.,
Christou, J. C.\& Probst, R. G. 1993, \apj, 406, 112

\bibitem[Feigelson\etal (2002)]{feigelson02} Feigelson, E. D.\etal 2002, ApJ, 574,  258

\bibitem[Garay \& Lizano (1999)]{gl99} Garay, G., \& Lizano, S. 1999, PASP, 111, 
1049

\bibitem[Garmire\etal (2000)]{garmire00} Garmire, G., Feigelson, E. D., Broos,
P., Hillenbrand, L. A., Pravdo, S. H., Tonwsley, L., \& Tsuboi, Y. 2000, AJ, 120,
1426

\bibitem[Genzel \& Stuzki (1989)]{genzel89} Genzel, R., \& Stuzki, J. 1981, \araa,
27, 41

\bibitem[Gezari (1992)]{dyg92} Gezari, D. Y. 1992, ApJ, 396, L43

\bibitem[Gezari\etal (1992)]{gezari92} Gezari, D., Folz, W.,
Woods, L., \& Varosi, F. 1992, PASP, 104, 191

\bibitem[Gezari\etal (1998)]{gezari98} Gezari, D. Y., Backman, D. \& Werner, M. W.
1998, ApJ, 509, 283

\bibitem[Greenhill\etal (1998)]{greenhill98} Greenhill, L. J., Gwinn, C. R., Schwartz, C., Moran,
J. M., Diamond, P. J.  1998, \nat, 396, 650  

\bibitem[Greenhill\etal (2003)]{greenhill03} Greenhill, L. J., Chandler, C. J., Reid, M. J., Diamond,
P. J., \& Moran, J. M.  2003, Proc. IAU Symp 221,  eds. M. Burton, R. Jayawardhana, \& T. Bourke
ASP Conference Series, in press.

\bibitem[Hillenbrand \& Hartmann (1998)]{hh98} Hillenbrand, L. A. \& Hartmann, L. W. 1998, ApJ, 492, 540

\bibitem[Hillenbrand \& Carpenter (2000)]{hc00} Hillenbrand, L. A., \& Carpenter, J. M. 2000, ApJ, 540, 236

\bibitem[Johnston\etal (1989)]{johnston89} Johnston, K. J., Migenes, V., \&
Norris, R. P. 1989, ApJ, 341, 847

\bibitem[Jones \& Puetter (1993)]{jp93} Jones, B., \& Puetter, R. 1993, Proc. SPIE, 1946, 610

\bibitem[Kroupa (2004)]{kroupa04} Kroupa, P. 2004, New Astr Rev, in press

\bibitem[Lonsdale\etal (1982)]{lonsdale82} Lonsdale, C. J., Becklin, E. E., Lee, 
T. J., \& Stewart, J. M. 1982, \aj, 87, 1819

\bibitem[Luhman\etal (2000)]{luhman00} Luhman, K. L.,\etal 2000, ApJ, 544, 1016
 
\bibitem[Mason\etal (2000)]{mason00} Mason, B. D.,\etal 2000, AJ, 120, 1120

\bibitem[Mathis (1990)]{mathis90} Mathis, J. S. 1990, ARAA, 28, 37

\bibitem[Menten \& Reid (1995)]{mr95} Menten, K. M., \& Reid, M. J. 1995, \apj, 
445, 157

\bibitem[Migenes\etal (1989)]{migenes89}Migenes, V., Johnston, K. J., Pauls, T. A., \& Wilson, T. L.
1989, ApJ, 347, 294
\
\bibitem[Monnier \& Millan-Gabet (2002)]{mmg02} Monnier, J. D. \& Millan-Gabet, R. 2002, ApJ, 579, 694

\bibitem[Plambeck\etal (1995)]{plambeck95} Plambeck, R. L., 
Wright, M. C. H., Mundy, L. G. \& Looney, L. W. 1995, ApJ, 455,
L189

\bibitem[Rouan\etal (2004)]{rouan04} Rouan, D.\etal 2004, A\&A, in prep.

\bibitem[Schultz\etal (1999)]{schultz99} Schultz\etal 1999, ApJ, 511, 282

\bibitem[Scoville\etal (1983)]{scoville83} Scoville, N., Kleinmann, S. G., Hall, D. N. B., \&
Ridgway, S. T. 1983, ApJ, 275, 201

\bibitem[Shuping\etal (2004)]{shuping04} Shuping, R. Y., Morris, M., \& Bally, J. 2004, AJ, submitted

\bibitem[Stolovy\etal (1998)]{stolovy98} Stolovy, S.,\etal 1998, ApJ, 492, L151

\bibitem[Testi\etal (1999)]{testi99} Testi, L., Palla, F., \& Natta, A.
1999, A\&A, 342, 515

\bibitem[Tuthill\etal (2001)]{tmd01} Tuthill, P. G., Monnier, J. D., \& Danchi, W. C. 2001, Nature, 409,
1012

\bibitem[Varosi \& Gezari (1993)]{vg93} Varosi, F. \& Gezari, D. Y. 1993,
``Astronomical Data Analysis Software and Systems II,''  PASP Conf Series, 52,
393

\bibitem[Wright\etal (1996)]{wright96} Wright, M. C. H., Plambeck, R.
L., Wilner, D. J. 1996, ApJ, 469, 216

\end{thebibliography}
\end{document}